\documentclass[a4paper,11pt]{article}
\usepackage{pos}
\title{One inch LaBr3:Ce detectors, with temperature control and improved
time resolution for low energy X-rays spectroscopy} 
\ShortTitle{One inch LaBr3:Ce detectors read by SiPM arrays}
\author*[a]{M. Bonesini}
\author[a]{R. Benocci}
\author[a]{R. Bertoni}
\author[b]{A. Abba}
\author[b]{F. Caponio}
\author[c]{A. Menegolli}
\author[c]{M.C. Prata}
\author[c]{M. Rossella}
\author[c]{R. Rossini}

\affiliation[a]{Sezione INFN Milano Bicocca,Dipartimento di Fisica G. Occhialini
, Dipartimento di Scienze dell' Ambiente e della Terra, 
Universit\'a Milano Bicocca,  Piazza Scienza 3, 20123 Milano, Italy}

\affiliation[b]{Nuclear Instruments srl, via Lecco 3, 22045 Lambrugo, Italy}
\affiliation[c]{Sezione INFN Pavia, Dipartimento di Fisica, Universit\'a 
di Pavia,  via A. Bassi 6, 27100 Pavia, Italy}

\emailAdd{Maurizio.Bonesini@mib.infn.it}

\abstract{
Large area  LaBr$_3$:Ce detectors with a SiPM array 
readout have been developed for
the FAMU experiment at RAL. The aim was to have a good energy resolution 
for low energy X-rays detection ($\sim 100$ keV) and good timing properties 
of the signal pulse. Sixteen 1" detectors and twelve
1/2" detectors have been presently 
installed in the FAMU experiment 
and are taking data since March 2023, at RIKEN RAL Port 1. }

\FullConference{The European Physical Society Conference on High Energy Physics (EPS-HEP2023)\\
 21-25 August 2023\\
Hamburg, Germany\\}

\begin{document}
\maketitle

\section{Introduction}
LaBr$_3$:Ce  crystals, with Photomultiplier (PMT) or SiPM readout, have
applications in many fields from medical imaging, to homeland
security  and gamma-ray astronomy \cite{intro}. A readout based
on SiPMs or SiPM arrays, instead of PMTs,
allows the construction of compact detector, that may
be used in external magnetic fields, avoiding the use of complicate shielding
\cite{mice}. Good energy resolutions: around $3 \%$  at the 
Cs$^{137}$  peak (661.7 keV), comparable with the ones obtained with
PMTs \cite{pmt},  
have been recently obtained for large detectors 
(area > 1 $cm^2$) with a SiPM array readout \cite{divita}, 
\cite{poleshchuck}, \cite{bonesini}. 
Our efforts aimed at the development of detectors with large area (initially
1/2") with both a good energy resolution, down to 100 keV, and 
good timing properties, as shown in
reference \cite{bonesini1}. The purpose was to contribute to the X-rays
detection system of the FAMU experiment \cite{famu} at the 
RIKEN-RAL muon facility.
The experimental target of FAMU is the precision measurement of the hyperfine
splitting (HFS) of the $\mu$p ground state and thus of the proton Zemach
radius, with impinging muons. 
\section{Detectors' construction}
\begin{figure}[hbt]
\begin{center}
\includegraphics[width=16.0 cm]{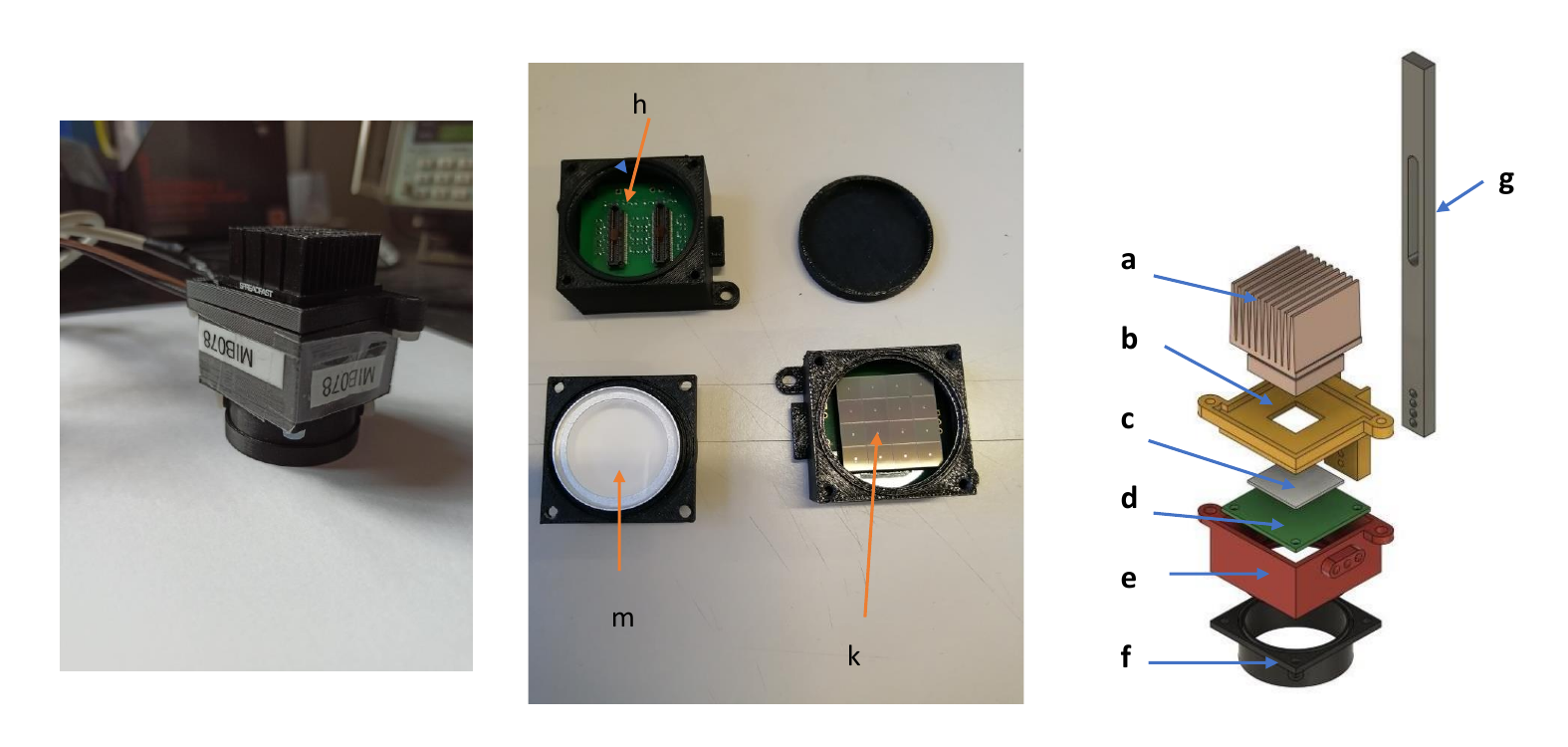}
\end{center}
\caption{Left panel: image of a complete detector. Middle panel:
images of some details of the crystal holder; h) with PCB
inside. The two SAMTEC connectors for SiPM array mounting are shown.
k) with mounted SiPM array; m) with crystal inside.   
Right panel: exploded view of a 1" detector. From top to bottom: 
a) heat dissipator, b) detector base; c) gap filler, d) PCB ,
e) PCB holder, f) crystal holder, g) support to hang a detector's base (b) 
from the LaBr$_3$:Ce  crown support.}
\label{fig-0}
\end{figure}

The developed detectors are based on 1" round LaBr$_3$:Ce 
crystals (0.5 " thick) read by 
Hamamatsu S14161-6050AS-04 square 1" SiPM array and are shown in figure
\ref{fig-0}.  

All the needed mechanics items  for the detector mounting were realized with a 
3D printer. Further details are reported in references \cite{bonesini1} and 
\cite {bonesini4}.
\begin{figure}[hbt]
\begin{center}
\includegraphics[width=10.5 cm]{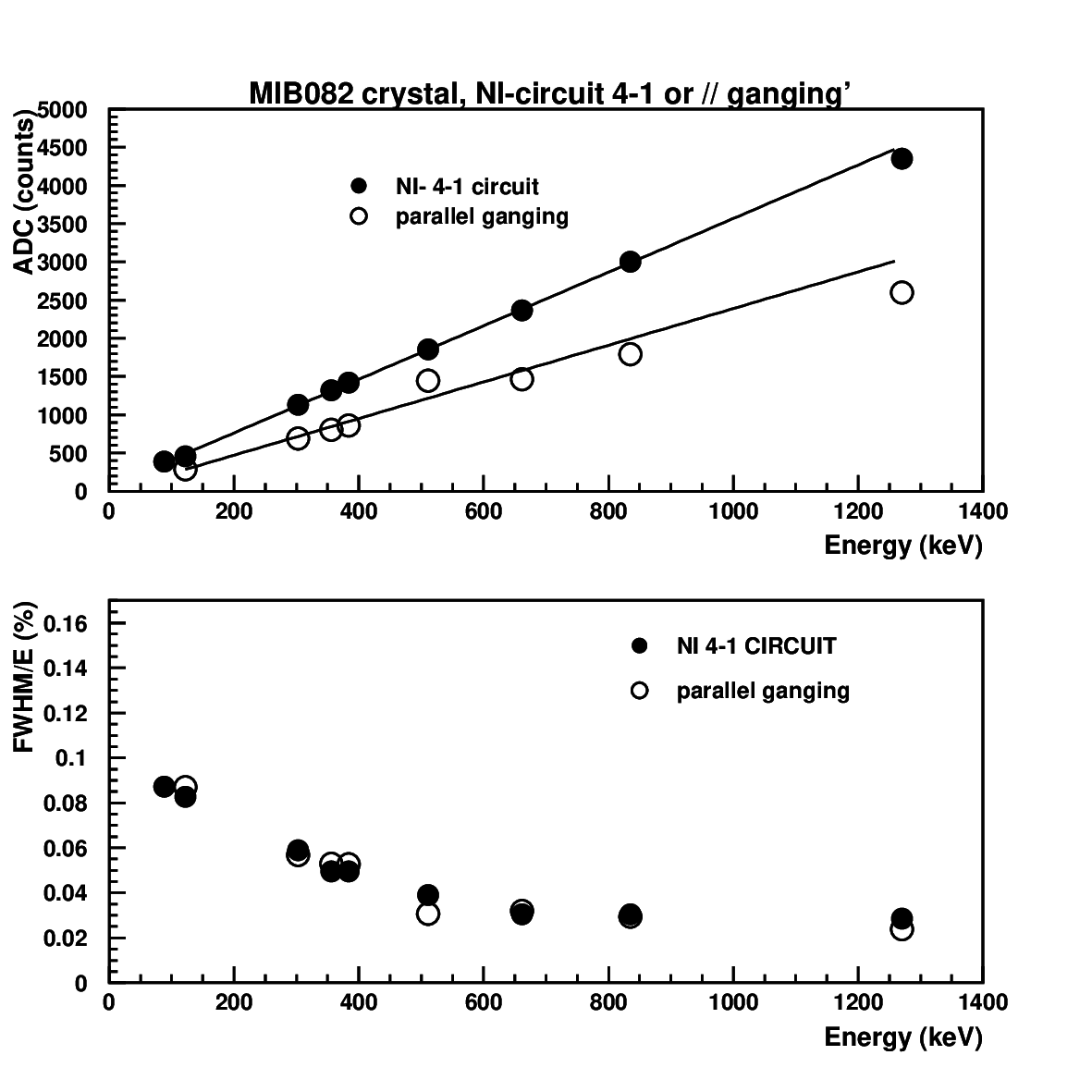}
\end{center}
	\caption{Linearity (top panel) and FWHM energy resolution (bottom
	panel) for a typical 1" detector both with parallel ganging and the
	new 4-1 circuit from Nuclear Instruments.}
        \label{fig-1}
\end{figure}
The ouput signals of four
nearby 6x6 mm$^2$ SiPM are grouped together and for them an individual 
pole-zero 
compensation and amplification is applied. Afterwords, 
the four signals of the sub-arrays (2x2 cells) 
are summed together and inverted, to give a positive output. All is realized 
with a custom circuit realized with Nuclear Instruments srl, see reference
\cite{bonesini5} for additional details. In this way a good energy resolution 
is  obtained and pulse timing 
(risetime/falltime) is reduced by a factor 2X. 
Figure \ref{fig-1} shows the FWHM energy resolution and the linearity for
a typical 1" detector, both with the standard parallel ganging for the 
16  cells of the used SiPM array and with the custom 4-1 circuit from Nuclear 
Instruments srl. No deterioration of FWHM energy resolution is evident. 
The well-known drift of the SiPM gain with temperature is
corrected online via a NIM custom module, based on  CAEN A7585D
power supply modules, with temperature feedback. 
In the range 10-35 $^{\circ}$C the effect on pulse height (P.H.) 
variation at the $^{137}$Cs photo-peak is reduced from 40 \% to 7 \%,
see references \cite{bonesini4}, \cite{bonesini2} for additional details.
\section{Experimental Results}
Laboratory tests were performed at INFN MIB with exempt sources from 
Spectrum Techniques, covering the energy range 80-1274.5 keV ($^{109}$Cd -
$^{22}$Na). Detectors under test were put inside a Memmert IPV30 climatic
chamber, with a  precision of $\pm 0.1 ^{\circ}$C on the temperature control.  
Detectors' signals were then fed into a CAEN VME V1730 FADC and
then acquired by the FAMU custom DAQ, via a CAEN V2718 interface. 
Output ROOT ntuples are then processed for data analysis. 

\begin{figure}[hbt]
\begin{center}
\includegraphics[width=8.0 cm]{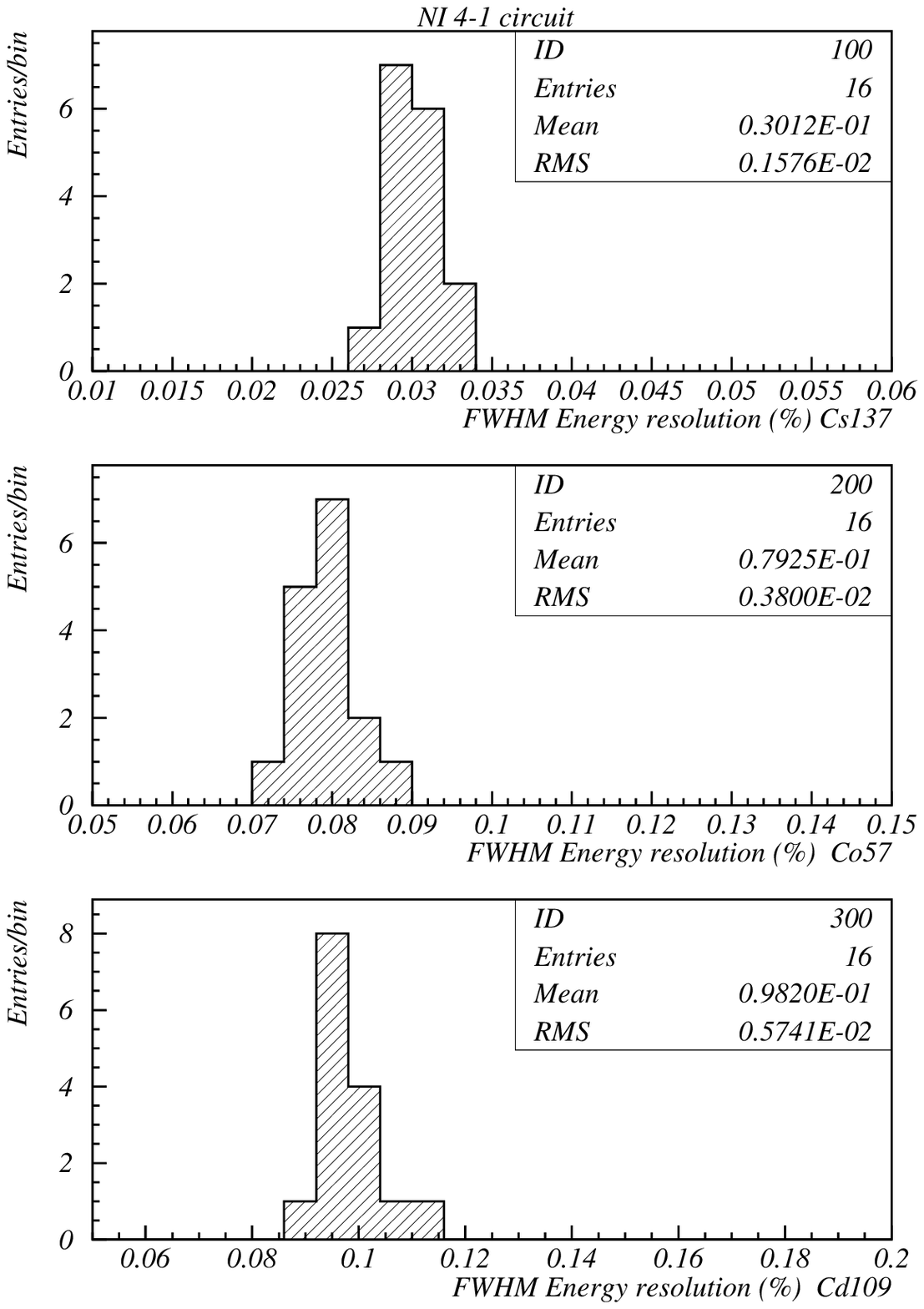}
\end{center}
	\caption{Distribution of FWHM energy resolution for the used 1" 
	detectors 
	at the $^{137}$Cs peak (top panel), at the $^{57}$Co peak (middle panel)
	and at the $^{109}$Cd peak (bottom panel). }
        \label{fig-2}
\end{figure}
\begin{figure}[hbt]
\begin{center}
\includegraphics[width=7.0 cm]{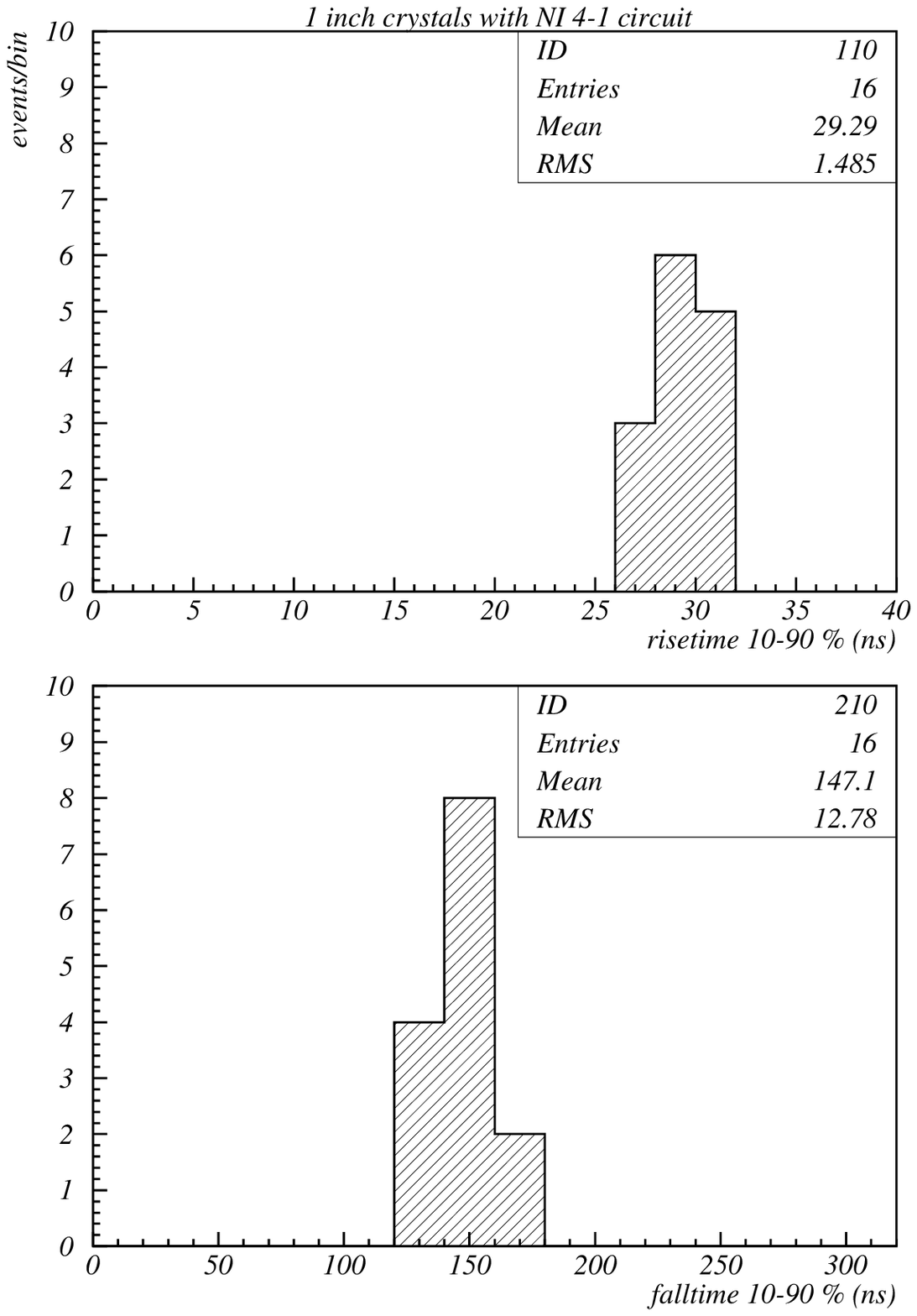}
\end{center}
	\caption{Distribution of the 10-90 $\%$ risetime-falltime for the
	used 1" detectors.}
        \label{fig-3}
\end{figure}

FWHM energy resolution (in $\%$) and 10-90 $\%$ risetime/falltime 
for the sixteen 1" detectors used in the FAMU experiment are shown 
in figure \ref{fig-2} and \ref{fig-3}. The main experimental challenge 
in the FAMU experiment is the detection of low-energy X-rays around 130 keV
separating a "delayed" signal component (after ~ 600 ns) from a prompt one.
The reported results: especially the $\sim 8 \%$ energy resolution at the
$^{57}$Co peak and the signal falltime ($\sim 150$ ns) show that 
the required
target may be reached \cite{bonesini1}, \cite{bonesini5}. 
Prelimary data taken at RIKEN RAL with 55 MeV muons
from March 2023 confirm this 
assumption. 
\subsection{Installation at RIKEN RAL}
\begin{figure}[hbt]
\begin{center}
\includegraphics[width=12.5 cm]{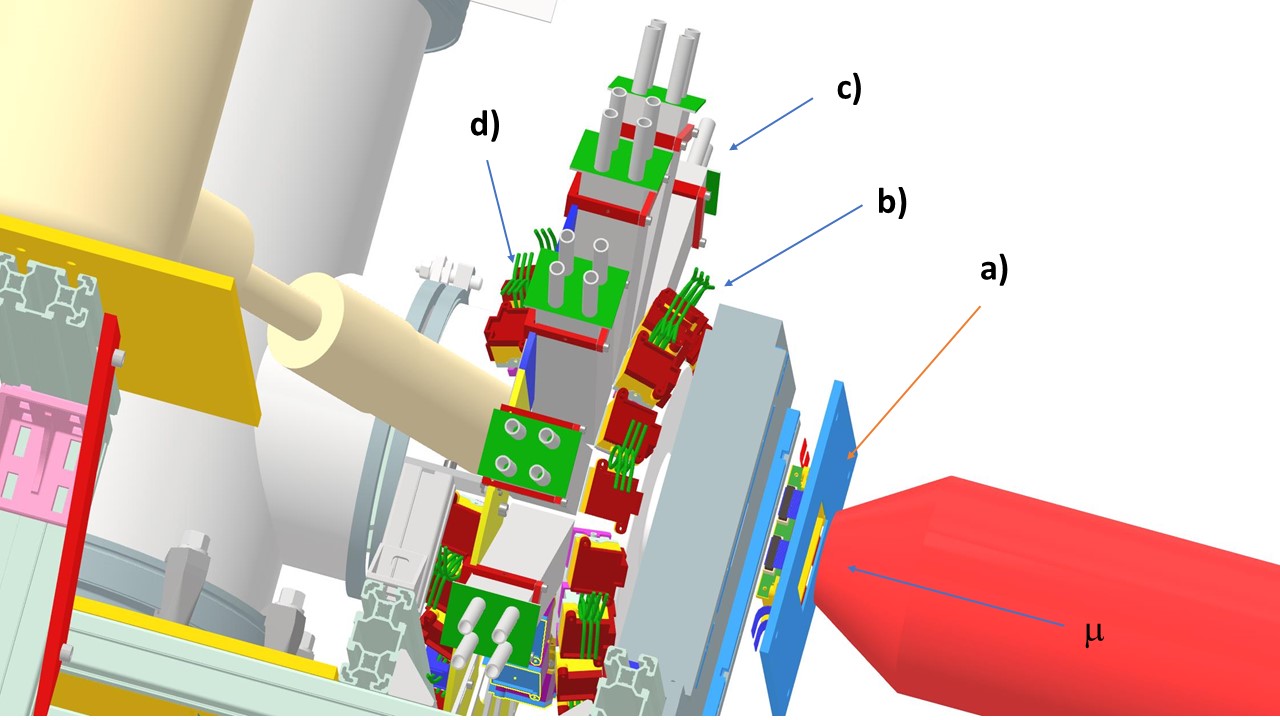}
\end{center}
        \caption{Drawing of the FAMU X-rays system, used in the 2023
	run, with (a) the beam
        hodoscope in front of a Pb collimator,
        (b) the upstream crown of 1" LaBr$_3$:Ce detectors;
        (c) the central crown of detectors: based on six old detectors 
	read by PMTs \cite{adam} and
	six detectors read by SiPM arrays (visible near the bottom). 
	All are of 1" size. 
        (d) the downstream crown of
        1/2" Ce:LaBr$_3$ detectors.}
        \label{ral}
\end{figure}

In the 2023 run configuration for the FAMU experiment at RIKEN-RAL,
sixteen 1"  and twelve 1/2"  detectors, with SiPM readout, were
installed at RAL in the X-ray detection system.
They are arranged in three crowns : upstream (ten 1" detectors), 
central (six 1" detectors) and downstream (twelve 1/2" detectors),
as shown in figure \ref{ral}. In addition six old 1" detectors with a PMT 
readout \cite{adam} are used in the central crown. 

With the 4-1 innovative circuit from  Nuclear Instruments srl, that divide
the readout of a 1" SiPM array into four, a good compromise in 
the optimisation
of energy resolution and pulse timing is obtained.
 The major drawback of this solution is the
increase of heat dissipation, due to the introduction
of several Texas Instruments OPA695 amplifiers: about 1 W. 
As the working environment at RIKEN-RAL 
is kept at constant temperature (20 $^{\circ}$C)
by air-conditioning, a simple passive heat dissipation is enough 
for proper operations.

The main characteristics of the Ce:LaBr$_3$ detectors used in the FAMU
experiment, as obtained  from laboratory measurements,
are resumed in table \ref{tab2}.
10-90 $\%$ risetime (falltime) are measured with a Cs$^{137}$ source, using
a Lecroy 1 GHZ scope. 
\begin{table}[htb]
\caption{FWHM energy resolution and timing characteristics of the Ce:LaBr$_3$
        detectors used in the FAMU experiment
        at RIKEN-RAL \label{tab2}}
\smallskip
\begin{tabular}{lccccc}
\hline
&\textbf{risetime (ns)} &  \textbf{falltime (ns)}     
& \textbf{resolution \%} & \textbf{resolution \% } & \textbf{resolution \%} \\
	&   &  & \@ Cs$^{137}$    &    \@ Co$^{57}$  & \@ Ce$^{109}$ \\ 
	&   &  &   (662 keV)    &     (122 keV)    & (88 keV)       \\
\hline
1/2" detectors & 42.8 $\pm$ 4.7 & 372.4 $\pm$ 17.4 & 3.27 $\pm$ 0.11 & 8.44 $\pm$ 0.63 &  10.63 $\pm$ 1.16 \\
	1" detectors & 29.3 $\pm$ 1.5 & 147.1 $\pm$ 12.8 & 3.01 $\pm$ 0.16 & 7.93 $\pm$ 0.38 & 9.82 $\pm$ 0.54 \\
\hline
\end{tabular}
\end{table}

The worse timing properties of the 1/2" detectors, as compared to the 1" ones,
are mainly due to the adoption of a standard parallel ganging instead of the
4-1 solution from Nuclear Instruments and probably to a different Ce
concentration.
Data were taken since March 2023 with a 55 MeV/c muon beam at  Port 1 at RIKEN RAL and a preliminary
analysis is under way. No major issues have been encountered up to now.
Preliminary results on the reconstructed  muonic Ag peak ($\sim 142$ keV) 
confirm our laboratory ones.

\section{Conclusions}
Good FWHM energy resolution is obtained with 1" LaBr$_3$:Ce crystals read by
Hamamatsu S14161-6050AS-04 SiPM arrays. Resolutions better than $3 \% (8 \%)$
are obtained at the Cs$^{137}$ (Co$^{57}$) peak in laboratory tests. 
The use of the innovative 4-1
circuit from Nuclear Instruments allowed a factor two reduction of signal
risetime (falltime), as respect to the conventional solution with parallel
ganging. Detectors are presently installed in the FAMU apparatus at RIKEN RAL.
\section*{Acknowledgements}
We would like to thank all members of the FAMU collaboration for
friendly discussions, in particular  Andrea Vacchi,  Ludovico
Tortora and  E. Mocchiutti.
We acknowledge also the help in detectors' mounting and the realization
of the test setup of Luca Pastori, Nuclear Instruments srl,
Roberto Gaigher, Stefano Banfi  and Giancarlo
Ceruti, INFN Milano Bicocca mechanics workshop and of  Maurizio Perego ,
INFN Milano Bicocca,
for SPICE simulation of the used electronic circuits.

\end{document}